\date{\today}

\documentclass[12pt,a4paper]{article}

\usepackage[affil-it]{authblk}
\usepackage{graphicx}
\usepackage{epstopdf} 
\usepackage{cite}
\usepackage{caption}
\usepackage{subcaption}
\usepackage[]{units}

\makeatletter
\renewcommand{\@maketitle}{%
  \begin{center}
    \textbf{\@title}
    \par
    \vspace{10pt}
    \textit{\@author}
      \end{center}
  \par
}
\makeatother

\title{Spectroscopy Apparatus for the Measurement of The Hyperfine Structure of Antihydrogen}

\author[1]{C.~Malbrunot\thanks{Present address: CERN, Gen\`eve 23, CH-1211, Switzerland. Electronic address: \texttt{chloe.m@cern.ch}}}
\author[1]{P.~Caradonna}
\author[1]{M.~Diermaier}
\author[1]{N.~Dilaver}
\author[1]{S.~Friedreich}
\author[1]{B.~Kolbinger}
\author[1]{S.~Lehner}
\author[2]{R.~Lundmark}
\author[1]{O.~Massiczek}
\author[3]{B.~Radics}
\author[1]{C.~Sauerzopf}
\author[1]{M.~Simon}
\author[1]{E.~Widmann}
\author[1]{M.~Wolf}
\author[1]{B.~W\"unschek}
\author[1]{J.~Zmeskal}
\affil[1]{Stefan Meyer Institute for Subatomic Physics, Austrian Academy of Sciences, Boltzmanngasse 3, 1090 Vienna, Austria}
\affil[2]{Department of Fundamental Physics, Chalmers University of Technology, SE-41296 G\"oteborg, Sweden}
\affil[3]{ RIKEN Advanced Science Institute Hirosawa, Wako, Saitama 351-0198, Japan}

\begin{document}

\maketitle

\begin{abstract}
The ASACUSA CUSP collaboration at the Antiproton Decelerator (AD) of CERN is planning to measure the ground-state hyperfine splitting of antihydrogen ($\bar{\textrm{H}}$) using an atomic spectroscopy beamline. We describe here the latest developments on the spectroscopy apparatus developed to be coupled to the $\bar{\textrm{H}}$ production setup (CUSP).

\end{abstract}

\section{Introduction}
\label{intro}
Antihydrogen is the simplest atom consisting entirely of antimatter. Since the atomic energy levels of its matter counterpart are very precisely measured quantities, a comparison of antihydrogen and hydrogen could provide one of the most sensitive tests of CPT symmetry \cite{Colladay, Bluhm}. In addition to providing a CPT test, the measurement of the ground state hyperfine splitting (GSHFS) at a relative precision of 10$^{-6}$, in combination with the precise measurement of the antiproton ($\bar{\textrm{p}}$) magnetic moment \cite{DiSiacca}, would probe the internal structure of the $\bar{\textrm{p}}$ through its electric and magnetic form factors appearing in the Zemach correction term to the GSHFS at the \unit[41]{ppm} level \cite{Friar}. 
\section{Method}
Measurements of the 1S-2S transition \cite{ATRAP, ALPHA} and GSHFS \cite{ALPHA2} of $\bar{\textrm{H}}$ in an atomic trap requires cold ($<$\unit[0.5]{K}) atoms.  The measurement of the GSHFS in a beam as proposed by the ASACUSA collaboration \cite{Widmann1, Widmann2} can work up to a beam temperature of \unit[100]{K} which eliminates the challenge of producing very cold $\bar{\textrm{H}}$. The principle of the experiment is based on the work of Rabi \cite{Rabi} and makes use of the splitting of the hyperfine levels in the presence of an external magnetic field. Atoms in the states (F,M$_F$)=(1,-1) and (F,M$_F$)=(1,0) are deflected into regions with lower fields and hence are called low field seekers (LFS), F being the total angular momentum quantum number on the antiproton and $-F\leq \textrm{M}_F\leq+F$. Those in the state (F,M$_F$)=(1,1) and (F,M$_F$)=(0,0) are high field seekers (HFS). An inhomogeneous magnetic field at the location of the $\bar{\textrm{H}}$ production can therefore focus LFS and defocus HFS producing the needed initial polarization of the beam \cite{Enomo}. A microwave field in a resonant cavity can drive the transition (either $\sigma$ or $\pi$, depending on the orientation of the radio-frequency (RF) field with respect to the static magnetic field) between LFS and HFS. A magnet following the cavity would defocus particles which have undergone the transition producing a dip  in the signal recorded by the $\bar{\textrm{H}}$ detector when the RF field is on resonance.
\section{Experimental Setup}
\begin{figure}[h!]
\centering
 \includegraphics[width=0.85\textwidth]{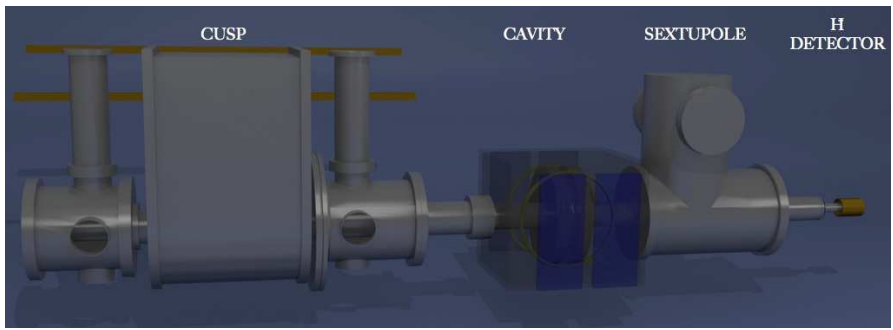}
\caption{Schematics of the beam apparatus for the measurement of the GSHFS of $\bar{\textrm{H}}$.}
\label{fig:apparatus}     
\end{figure}

The spectroscopy setup consists, after a source (CUSP) of partially polarized $\bar{\textrm{H}}$ atoms \cite{Mohri}, of an RF spin-flip cavity tuned to the hyperfine transition frequency, a \unit[3.5]{T} superconducting sextupole magnet which serves as spin analyzer and a detector to measure the products of the $\bar{\textrm{H}}$ annihilation  \cite{Widmann}, see Fig. \ref{fig:apparatus}. 

\subsection{Current Status}
\begin{figure}[h!]
\centering
\begin{minipage}{.5\textwidth}
  \centering
 \parbox{5.7cm}{
  \includegraphics[width=.85\linewidth]{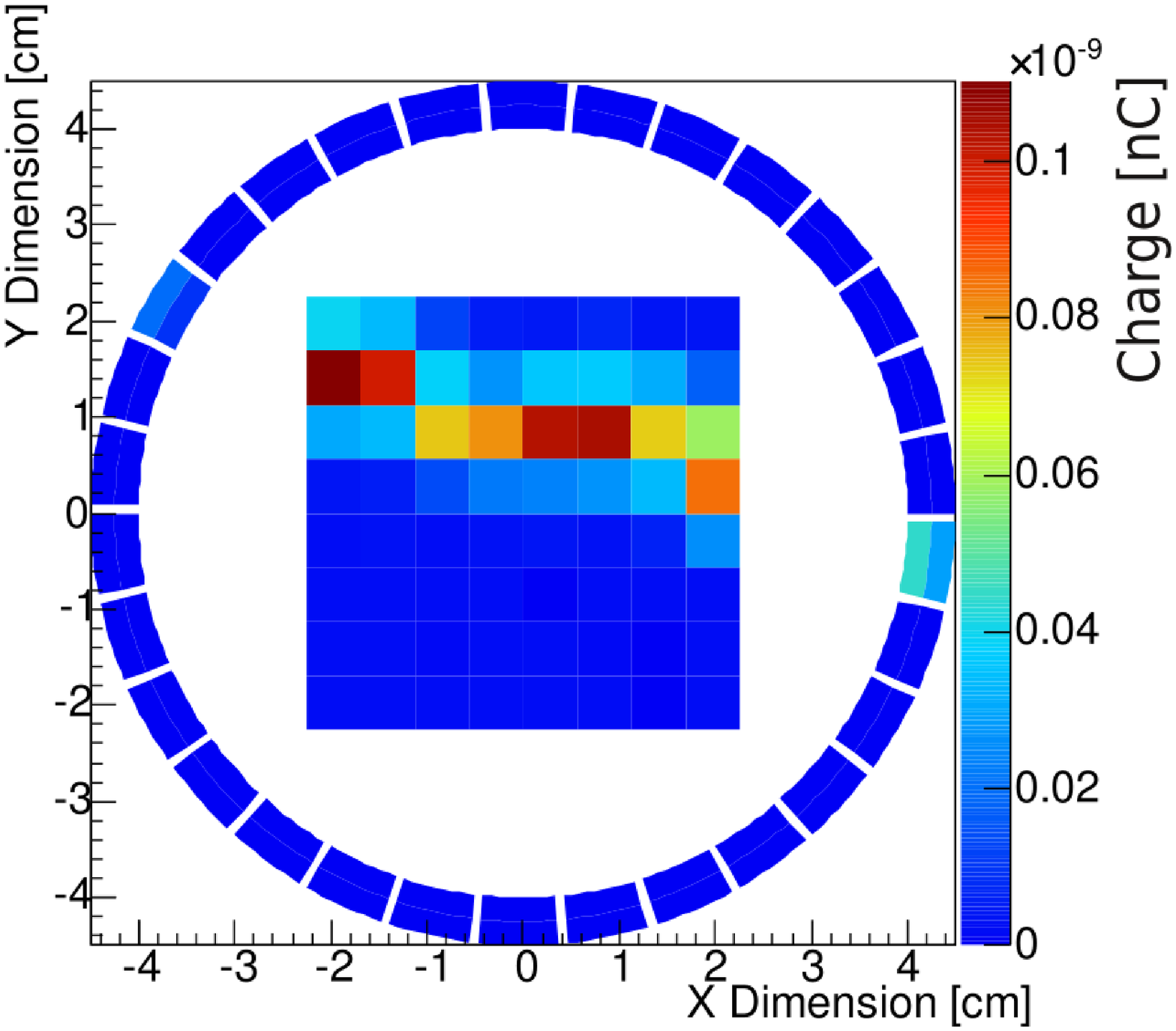}
  \caption{Online display of a cosmic signal recorded by the $\bar{\textrm{H}}$ detector. The $\bar{\textrm{H}}$  annihilate onto the central segmented detector. The hodoscope being read from both sides (the rectangular-shaped blocks separated in two represent the upstream and downstream (in z direction) parts of the hodoscope bars) can provide information on the z-position of the particle hit.}
  \label{fig:cosmics}}
\end{minipage}%
\begin{minipage}{.5\textwidth}
  \centering
   \parbox{5.7cm}{
  \includegraphics[width=.85\linewidth]{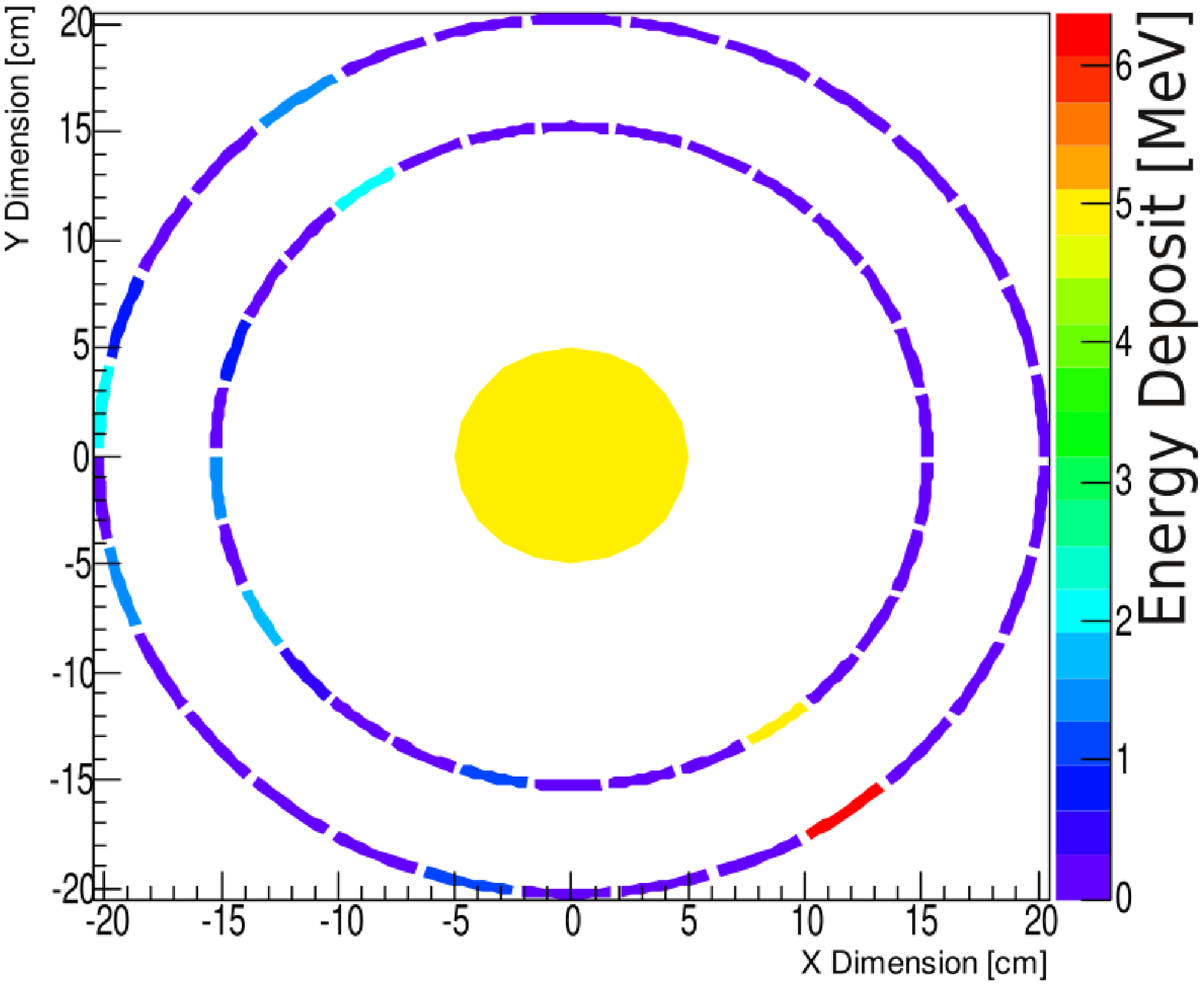}
  \caption{Geant4 simulation of an $\bar{\textrm{H}}$ annihilation onto a central calorimeter. The products of the annihilation (in this case three charged pions) are detected by two concentric hodoscope detectors read on both sides. The combination of energy deposit information and vertex reconstruction would improve the signal to noise ratio.}
  \label{fig:hannili}}
\end{minipage}
\end{figure}

The superconducting sextupole, the cavity as well as the detector consisting of a central segmented plastic scintillator surrounded by 30 scintillator bars assembled in a hodoscope geometry were developed and built in preparation for the 2012 data taking. Cosmic muons passing through the detector (see Fig.~\ref{fig:cosmics}) were recorded and helped evaluate the detector performance in identifying the annihilation products of the $\bar{\textrm{H}}$ against suppressing the background coming from cosmic rays and annihilation of $\bar{\textrm{H}}$ and $\bar{\textrm{p}}$ upstream of the detector.  Following this evaluation a series of improvements was undertaken. 
\\
The amplifier electronic of the silicon photomultipliers which read out the hodoscope bars was modified to improve the time resolution by a factor of $\sim$5 to become \unit[600]{ps} (FWHM). This improvement will provide a much better position resolution and cosmic discrimination in the hodoscope. A diagnostic detector to be placed after the CUSP, consisting of a central calorimeter surrounded by two layers of hodoscope, is currently under development, see Fig. \ref{fig:hannili}. This configuration would allow vertex reconstruction and provide additional energy deposit information which would improve the signal to noise (S/N) ratio.
\subsection{Further Developments}
The homogeneity in the cavity volume of the static field produced by the Helmholtz coils in the presence of the earth magnetic field has been measured to be $\sim$1$\%$ RMS which is, according to simulation \cite{Berti}, sufficient for the measurement of the $\sigma$ transition. Currently, the cavity and the pair of coils are surrounded by two layers of mu-metal shielding which is not sufficient to shield the cavity  against the CUSP stray field  at the level required for the measurement of the $\pi_1$ transition. A new shielding has to be designed in order to also accommodate for the additional coils needed for the simultaneous measurement of the  $\pi_1$ and $\sigma_1$ transitions. 

\section{Simulation}
\begin{figure}
\centering
 \includegraphics[width=1\textwidth]{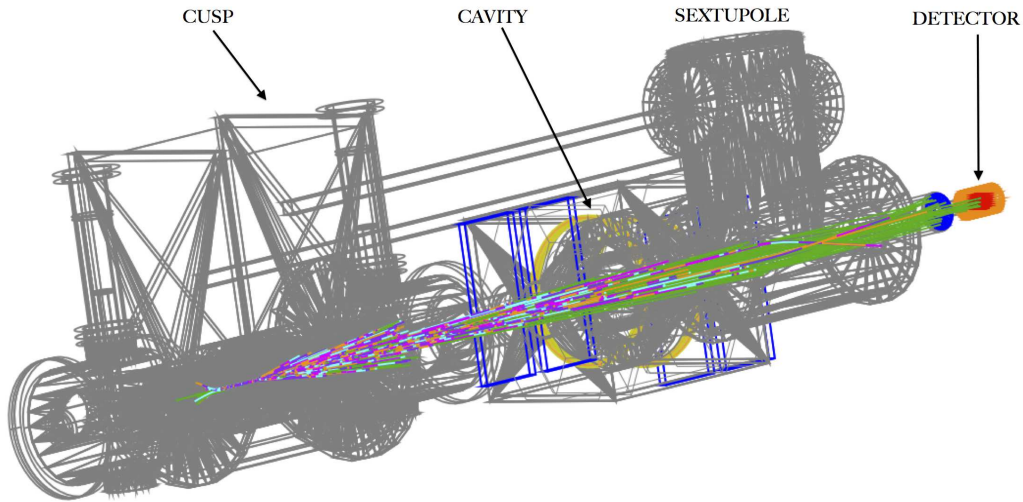}
\caption{Trajectories of $\bar{\textrm{H}}$  atoms in the beamline as simulated by a modified version of Geant4. The CUSP trap, the cavity, and sextupole are shown in gray, while the blue color denotes detectors to measure annihilation products at the cavity and before the $\bar{\textrm{H}}$ detector \cite{Widmann}. The red square is the current central antihydrogen detector surrounded by the hodoscope in orange. The effect on the trajectory and on the S/N ratio of the $\bar{\textrm{H}}$ radiative decay assuming a certain initial quantum state distribution and temperature at their production position in the CUSP has been simulated. The principal quantum state of the atoms is visualized by different colors along the trajectory. The green line indicates ground state LFS $\bar{\textrm{H}}$.}
\label{fig:G4}     
\end{figure}

A detailed geometry of the apparatus has been implemented in Geant4, see Fig. \ref{fig:G4}. Version 4.9 of Geant4 has been modified to add tracking of neutral atoms in a magnetic field allowing the simulation of the $\bar{\textrm{H}}$ trajectories in the CUSP, cavity and sextupole fields.
Even though the presented experiment can be performed with an $\bar{\textrm{H}}$ beam temperature up to \unit[100]{K} colder $\bar{\textrm{H}}$ is preferable for several reasons. First, colder atoms in the CUSP will provide a better initial polarization improving the S/N ratio. Secondly, the atoms need to be in ground state when they reach the cavity. However, the assumed dominant $\bar{\textrm{H}}$ formation process in the CUSP is the three-body recombination which might preferentially form atoms in high Rydberg states. Therefore, slow atoms would have more time to decay to the ground state before reaching the cavity.  Finally, the precision of the GSHFS spectroscopy is limited by the time of flight of the atoms inside the cavity. Colder atoms would have a longer interaction time with the RF field and therefore provide a more precise measurement.
For these reasons it is crucial to have a precise understanding of the impact of the beam parameters on the spectroscopy precision. For this purpose, we are implementing radiative decay of  $\bar{\textrm{H}}$ in our Geant4 simulation which can be visualized in Fig. \ref{fig:G4}. Indeed, as the $\bar{\textrm{H}}$ decays, the force felt by the atom in the magnetic field is changing due to the change of its magnetic moment, which affects the trajectory of the atom in the field and in turn modifies its state distribution.\\
Further work is also being done to include in the simulation the Majorana spontaneous deexitation \cite{Majorana} in the absence of a magnetic field or due to sudden field intensity change. This effect would decrease the beam polarization and affect the S/N ratio.
\section{Hydrogen Beamline}
\begin{figure}
\centering
 \includegraphics[width=0.9\textwidth]{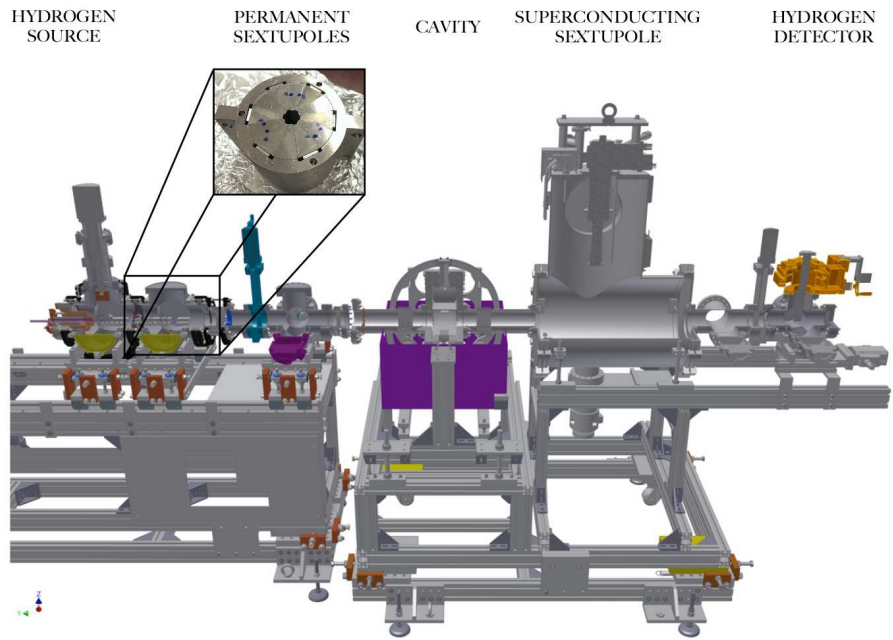}
\caption{Cut-through view of the H beamline. The H atoms enter the apparatus on the left of the figure and  the chamber containing the permanent sextupoles (see inset) which provide the initial beam polarisation.}
\label{fig:hbeam}     
\end{figure}

During the long LHC shutdown (end of 2012 to mid-2014), there will be no antiprotons available at the AD. We have developed a hydrogen (H) beam to further test the spectroscopy beamline in preparation for the next beam time in 2014, see Fig.\ref{fig:hbeam}. The atomic hydrogen is produced by dissociating highly purified molecular hydrogen in a  2.45 GHz microwave resonator. The beam is then guided to a set of two permanent sextupoles which create the initial polarization and velocity selection. Simulation have shown that two sextupoles with an inner radius of 10~mm and a maximum field of 1.3~T can create a polarization of more than 90\%. The beam is then pulsed which enables time-of-flight measurements for beam characterization. After passing the cavity and the superconducting sextupole, the H beam is detected by a quadrupole mass spectrometer in conjunction with a lock-in amplifier.
\section{Conclusion and Outlook}
The spectroscopy apparatus required for the measurement of the hyperfine spectroscopy of $\bar{\textrm{H}}$ in a beam has been built. Tests of the detector have been performed with cosmic rays and improvements in the design are currently in progress. 
Extensive simulations of the complete apparatus have been performed in order to assess the influence of the beam parameters (e.g. initial quantum state distribution, temperature of the beam, $\bar{\textrm{H}}$ production position in the CUSP) on the achievable measurement precision.\\
A  H beam apparatus is currently being assembled and characterized. It will be coupled to the cavity and superconducting sextupole in order to characterize the apparatus and evaluate the systematics.  First data-taking with the H beam is planned for fall 2013.\\

\section*{Acknowledgements}
The authors would like to thank H.~Knudsen for providing  the hydrogen source. R.L., C.M. and B.R. acknowledge useful discussions with E. Stambulchik, and M. F. Gu for the Flexible Atomic Code. This work is supported by the European Research Council grant no. 291242-HBAR-HFS and the Austrian Ministry for Science and Research.

\end{document}